\documentclass[aps,prl,reprint,superscriptaddress]{revtex4-2}
\usepackage{graphicx}
\usepackage{rotating}
\usepackage{amssymb}
\usepackage{amsmath}
\usepackage{epsfig}
\usepackage[normalem]{ulem}
\usepackage{bm}
\usepackage{color}
\usepackage{subfigure}

\begin{document}

\title{Breakdown of stochastic resonance in complex networks}% Force line breaks with \\

\author{Jonah E. Friederich}
\affiliation{Institut für Theoretische Physik, Technische Universität Berlin, Hardenbergstraße 36, 10623 Berlin, Germany}

\author{Everton S. Medeiros}
\email{es.medeiros@unesp.br}
\affiliation{S\~ao Paulo State University (UNESP), Institute of Geosciences and Exact Sciences, Avenida 24A 1515, 13506-900, Rio Claro, S\~ao Paulo, Brazil}

\author{Sabine H. L. Klapp}
\affiliation{Institut für Theoretische Physik, Technische Universität Berlin, Hardenbergstraße 36, 10623 Berlin, Germany}

\author{Anna Zakharova}
\affiliation{Bernstein Center for Computational Neuroscience, Humboldt-Universität zu Berlin, Philippstraße 13, 10115 Berlin, Germany}

\date{\today}% It is always \today, today,

\begin{abstract}
In networked systems, stochastic resonance occurs as a collective phenomenon where the entire stochastic network resonates with a weak applied periodic signal. Beyond the interplay among the network coupling, the amplitude of the external periodic signal, and the intensity of stochastic fluctuations, the maintenance of stochastic resonance also crucially depends on the resonance capacity of each oscillator composing the network. This scenario raises the question: Can local defects in the ability of oscillators to resonate break down the stochastic resonance phenomenon in the entire network? Here, we investigate this possibility in complex networks of prototypical bistable oscillators in a double-well potential. We test the sustainability of stochastic resonance by considering a fraction of network oscillators with nonresonant potential landscapes. We find that the number of nonresonant oscillators depends nonlinearly on their dissimilarity from the rest of the network oscillators. In addition, we unravel the role of the network topology and coupling strength in maintaining, or suppressing, the stochastic resonance for different noise levels and number of nonresonant oscillators. Finally, we obtain a low-dimensional deterministic model confirming the results observed for the networks.
\end{abstract}

\maketitle

\section{Introduction}

The phenomenon of stochastic resonance (SR) construes a mechanism in which noisy random fluctuations act constructively to enhance the intensity of weak periodic excitations in nonlinear systems. Since the theoretical proposition of SR by Benzi et. al. in 1981 in an attempt to explain the observed regularity of Earth's ice ages \cite{Benzi1981}, the concept of SR has become multidisciplinary. The literature supporting this phenomenon involves theoretical, experimental and observational studies ranging from climate sciences \cite{Benzi1982,Ganopolski2002,Das2020} and neuronal dynamics \cite{Douglass1993,Gluckman1996} to lasers \cite{McNamara1988,Barbay2000} and electronic circuits \cite{Anishchenko1992,Anishchenko1994,Mantegna1994}. For reviews on SR, see Refs. \cite{Gammaitoni1998,Wellens2003,Mcdonnell2009}.

In networked systems, the onset of SR has been observed from various perspectives. For instance, spectral amplification of periodic inputs has been verified in one- and two-dimensional arrays of nonlinear oscillators \cite{Lindner1995,Sungar2000}. For oscillators networked via complex topologies, it has been found that the presence of long-range connections favors SR in the same fashion as synchronization among the oscillators \cite{Gao2001}. Later on, it was demonstrated that the typical signatures of SR are induced by diversity in coupled oscillators \cite{Tessone2006}. Subsequently, the minimum intensity of the periodic inputs for the onset of SR has been investigated in various networks by applying the periodic signal locally to only one oscillator \cite{Perc2008,Perc2008b,Ozer2009}. Moreover, the ubiquity of SR in networked systems been proven in variety of studies addressing different topological and dynamical aspects \cite{Wu2011,Wang2012,Semenov2016,Yamakou2020,Calim2021,Bai2022,Semenov2022}.

While the literature emphasizes the onset of stochastic resonance (SR) in a variety of contexts, to the best of our knowledge, no study has directly addressed the stability of SR in networks subject to local failures in the ability of units to resonate. Therefore, given the ubiquity and applicability of SR in networked systems, investigating the persistence of this phenomenon against adversities is crucial for ensuring the reliability of applications that depend on SR.

Here, we investigate the robustness of stochastically resonant networks by introducing failing units. Specifically, we consider networks composed of overdamped oscillators with a double-well potential, subject to periodic forcing and noise. Initially, we set the system parameters to ensure the occurrence of SR, defining a homogeneous resonant network. We then replace the potential of a subset of oscillators with non-resonant ones, featuring higher potential barriers. This procedure constitutes a perturbation to the SR phenomenon in our network. By varying the number of non-resonant oscillators and their level of dissimilarity with the resonant ones, we establish a robustness landscape for SR. By analyzing different parameter sets, we uncover the weakening and the thresholds at which SR globally fails. We then compare the breakdown scenarios among three coupling structures: regular, random, and scale-free. Finally, we apply an averaging procedure to obtain a low-dimensional deterministic approximation of our heterogeneous network, confirming our results.

\section{\label{sec:level1}The occurrence of stochastic resonance}

We consider networks composed of $N$ noisy bistable overdamped oscillators with motion given by the following Langevin equations \cite{Lindner1995,Gao2001}:
\begin{eqnarray}
\nonumber
\dot{x}_i &=& - \frac{\partial V_i(x_i,t)}{\partial x_i} + c \sum_{j=1}^N M_{ij} (x_j - x_i) + A \cos(\omega t) \\
 &+& D \xi_i(t),
\label{network_dgl_homo}
\end{eqnarray}
with $i=1, \dots, N$. The system size is fixed at $N=100$ oscillators. The potential $V_i(x_i,t)$ is a quartic function given by:
\begin{eqnarray}
V_i(x_i,t) = -\frac{1}{2} a_i x_i^2 + \frac{1}{4} b x_i^4,
\label{Potential1}
\end{eqnarray}
with $a_i,b\ \textgreater \ 0$. The coefficients $a_i$ of this potential introduce heterogeneities in our network corresponding to the failing oscillators. The parameter $b$ is kept fixed at $1.47$ throughout the study. The network coupling function is linear, representing a diffusive process between oscillators $i$ and $j$. The intensity of the coupling is controlled by the parameter $c$, while the coupling structure is specified by the symmetric matrix $\textbf{M}$. The entries of $\textbf{M}$ take the values $M_{ij} = 1$ if two oscillators $i$ and $j$ are coupled to each other, otherwise, $M_{ij} = 0$. Throughout this study, the matrix $\textbf{M}$ prescribes network topologies ranging from regular to random and scale-free \cite{Watts1998,Barabasi1999}. The system is subject to an external periodic forcing with amplitude specified by $A=1.30$ and frequency $\omega=2\pi \times 0.12$. The stochastic fluctuations $\xi_i(t)$ affecting each oscillator independently are given by a Gaussian white noise, specified as $\langle \xi_i(t) \xi_j(t- \tau) \rangle = \delta_{ij}\delta(t-\tau)$. The noise intensity $D$ of this stochastic process is uniform across the oscillators. We numerically solve Eq~\ref{network_dgl_homo} by implementing the Euler-Heun method \cite{euler_heun} with time step $dt=0.01$. For all simulations, we analyze the trajectories over a time interval of $T=300$, which includes several cycles of inter-well oscillations.

The onset of stochastic resonance is signaled by various spectral analysis tools \cite{Gammaitoni1998,Lucarini2019}. Here, we employ the spectral amplification $\eta$ to detect a peak value of the power spectrum at the forcing frequency $\omega$ for an applied noise intensity $D$. The spectral amplification is obtained by \cite{Wu2011, Tessone2006}:
\begin{eqnarray}
\eta = \frac{4}{A^2} | \langle e^{i \omega t} X(D,t) \rangle_t|^2,
\label{network_eta}
\end{eqnarray}
where $\quad X(t) = \frac{1}{N} \sum_{i=1}^{N} x_i(t)$ is a spatial mean. The operation $\langle \dots  \rangle_t = \frac{1}{T} \int_0^T f(t) dt$ denotes the time average.

\begin{figure}[!htp]
\centering
\includegraphics[width=85mm]{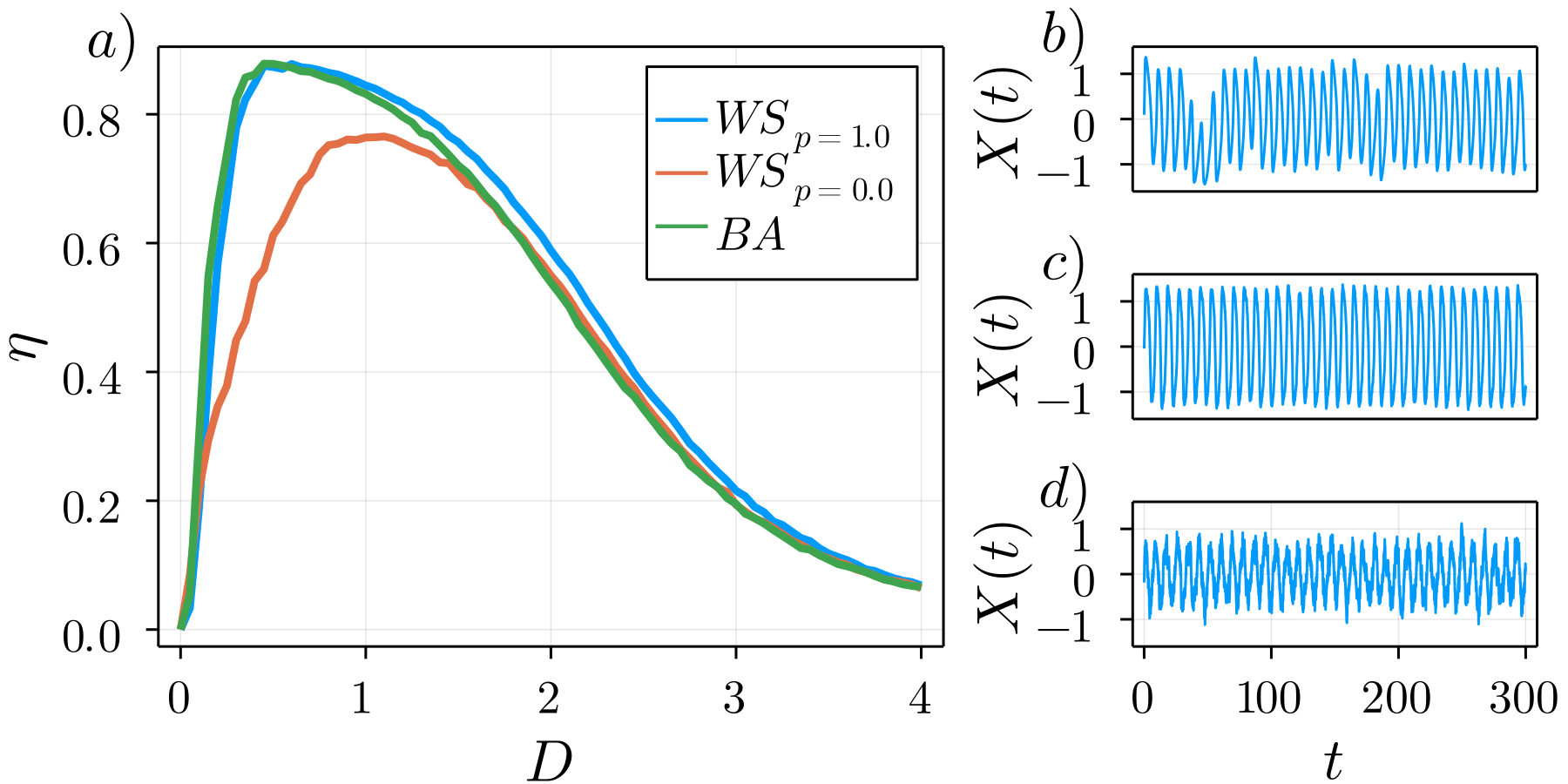}% Here is how to import EPS art
\caption{(a) Spectral amplification $\eta$ as a function of the noise intensity $D$ for three different topologies, $WS_{p=0.0}$ (orange), $WS_{p=1.0}$ (blue) and $BA_{m=2}$ topology (green). (b)-(d) For the $WS_{p=1.0}$ topology, we show trajectories of the spatial mean $X(t)$ for different noise intensities, $D=0.25$, $D=1.0$, and $D=3.5$, respectively. The coupling intensity is fixed at $c=1.0$.}
\label{figure1}
\end{figure}

We first illustrate the onset of stochastic resonance by obtaining $\eta$ for different values of $D$ in a system with homogeneous potential, i.e., $a_i=a$. The parameter $a$ is fixed at $a=2.11$ along our entire text. In Fig. \ref{figure1}(a), $\eta$ is calculated for $10$ different initial conditions (ICs) for each value of $D$. The different curves represent the various considered network topologies. The red curve corresponds to a regular topology generated using the Watts-Strogatz (WS) algorithm with rewiring probability $p=0.0$ and $k=4$ initial number nearest neighbors ($WS_{p=0.0}$). The blue curve stands for a random topology generated using the WS algorithm with $p=1.0$ and $k=4$ ($WS_{p=1.0}$). The green curve represents a scale-free topology generated using the Barab\'{a}si-Albert (BA) algorithm adding $m=2$ links at each step, ($BS_{m=2}$). Since stochastic resonance occurs in the inter-well dynamics, $\eta$ is obtained by neglecting the intra-well movement of the oscillators and using a two-state approximation $X(t) \rightarrow sgn(X(t)) = \pm 1$ \cite{Gammaitoni1998,Gao2001,Lindner1995}.

In Fig. \ref{figure1}(a), we observe well-pronounced maximum values of $\eta$ for all three topologies. This indicates the existence of optimum values of $D$ for the occurrence of regular inter-well motion at the forcing frequency $\omega$. Interestingly, networks with complex topologies, $WS_{p=1.0}$ and BA, exhibit higher $\eta$ for smaller values of the noise intensity $D$. The presence of long-range connections in these networks facilitates signal transmission across the network, thereby enhancing the onset of SR \cite{Gao2001}. The SR phenomenon can also be visualized in the time evolution of these networks. In Figs. \ref{figure1}(b)-\ref{figure1}(d), we present space-time diagrams illustrating the time evolution of the random network ($WS_{p=1.0}$) for three different values of $D$. For $D=0.25$, the spatiotemporal dynamics in Fig. \ref{figure1}(b) show an irregular inter-well motion as a result of subthreshold noise intensity. Conversely, for $D=1.0$ in Fig. \ref{figure1}(c), we observe regular motion between the potential wells occurring at the frequency of the external forcing $\omega$. In Fig. \ref{figure1}(d), we show the spatiotemporal dynamics for $D=3.5$, where the network dynamics become irregular again.

\section{Failing network oscillators and stochastic resonance breakdown}

We now investigate the robustness of stochastic resonance in the presence of structural defects, specifically subsets of oscillators that fail to resonate. To do this, we assign different values to the potential $V_i$  of each oscillator $i=1, \dots, N$.

\begin{figure}[!htp]
\centering
\includegraphics[width=85mm]{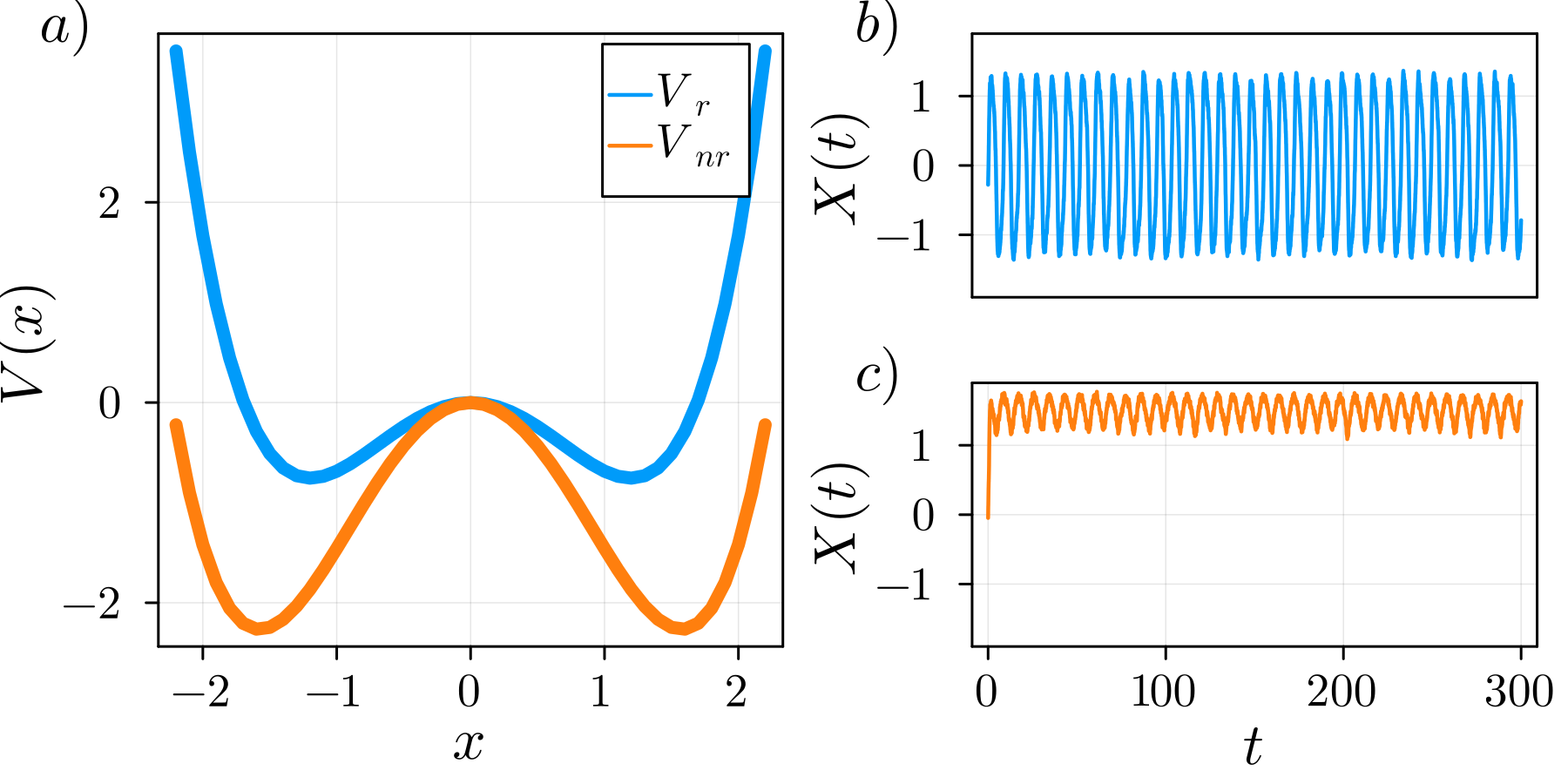}% Here is how to import EPS art
\caption{(a) The blue curve shows a resonant potential $V_{r}$, while the orange curve depicts a non-resonant one $V_{nr}$. (b) Time evolution of the mean variable $X(t)$ of a random network $WS_{p=1.0}$ solely composed of units with resonant potentials. (c) Corresponding time evolution of $X(t)$ for a $WS_{p=1.0}$ network composed exclusively of units with non-resonant potentials. The coupling and the noise intensity are $c=1.0$ and $D=1.0$, respectively.}
\label{figure2}
\end{figure}

Specifically, we divide $V_i$ into two domains based on the value of the potential coefficients $a_i$ of a subset $N_{nr} \in [0,N]$ of the network units that fail to resonate: i) For $i \leq N_{nr}$, the potential coefficients $a_i$ are set to $a_i=a_{nr}$, corresponding to a non-resonant potential $V_{nr}$, where the trajectories do not perform inter-well dynamics. The red curve in Fig. \ref{figure2}(a) depicts the potential landscape $V_{nr}$, and the time evolution of the network (Eq. (\ref{network_dgl_homo})) composed exclusively of non-resonant potential is shown in Fig. \ref{figure2}(c). ii) For $i>N_{nr}$, the potential coefficients $a_i$ are set to $a_i=a$, already discussed on the previous section, yielding a resonant potential $V_{r}$. The blue curve in Fig. \ref{figure2}(a) depicts the potential landscape $V_r$, and the corresponding time evolution of the network is shown in Fig. \ref{figure2}(b) (same as Fig. \ref{figure1}(c)). Naturally, the prescribed subset of units $N_{nr}$ corresponds to the number of units with non-resonant potentials in our network.

As illustrated in Fig. \ref{figure2}, the crucial feature for the establishment of inter-well dynamics in a homogeneous network (no failing oscillators) is the respective potential depth defined as $\Delta V_{i} = a_i^2/4b$. To control the level of dissimilarity between the failing oscillators the normal ones in our network, we introduce the parameter $\alpha_V$, which is the ratio between the non-resonant and resonant potential depths:
\begin{equation}
     \alpha_V \equiv \dfrac{\Delta V_{nr}}{\Delta V_{r}} \quad \Leftrightarrow \quad a_{nr} =  a \sqrt{\alpha_V}.
    \label{potratio}
\end{equation}
The parameter $\alpha_V$ prescribes the potential dissimilarity between the two network domains, $\alpha_V=1.0$ corresponds to the network with no failing oscillators. We regard the parameter $\alpha_V$ as a dissimilarity index. Hence, by selecting $\alpha_V$ and the number of units $N_{nr}$ with non-resonant potential, we can completely specify the failing conditions of the network in Eq. (\ref{network_dgl_homo}).

\begin{figure}[!htp]
\centering
\includegraphics[width=85mm]{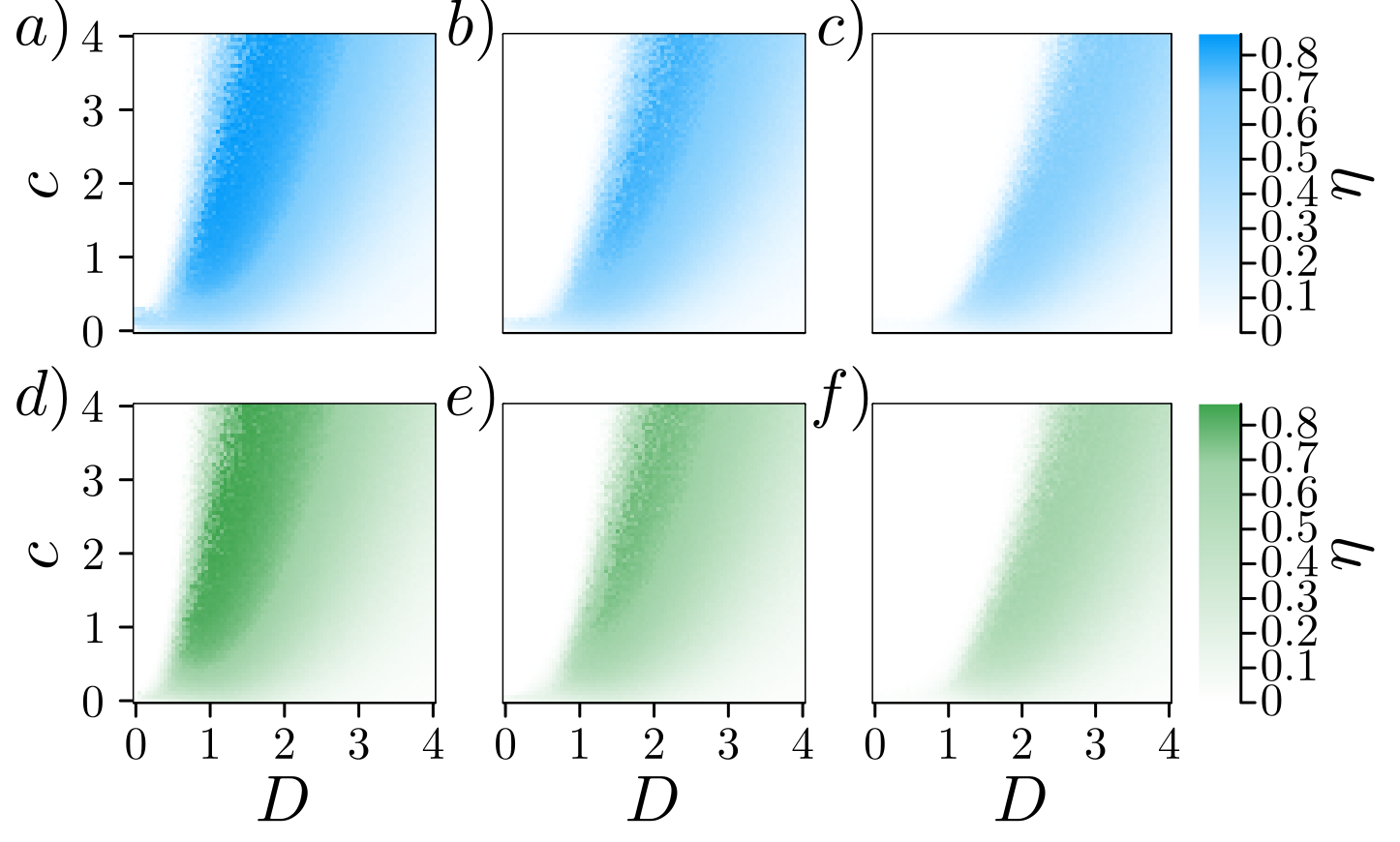}% Here is how to import EPS art
\caption{Spectral amplification $\eta$ (color code) as a function of the noise intensity $D$ and the coupling strength $c$. From left to right, the number of non-resonant units are $N_{nr}=10$, $N_{nr}=30$, and $N_{nr}=70$, respectively.  (a)-(c) For the $WS_{p=1.0}$ topology. (d)-(f) For the $BA_{m=2}$ the topology. The dissimilarity index is fixed at $\alpha_V=3.0$.}
\label{figure3}
\end{figure}

Following these developments, we now investigate the onset of stochastic resonance in networks containing various numbers of non-resonant units. The first step is to uncover the relationship between two important drivers of stochastic resonance in networks, namely, the coupling $c$ and the noise intensity $D$. To address such interplay, in Fig. \ref{figure3}, we obtain the spectral amplification $\eta$ for different values of $c$ and $D$ forming a two-dimensional parameter space. For a random network ($WS_{p=1.0}$), in Figs. \ref{figure3}(a)-\ref{figure3}(c), we fix the dissimilarity index at $\alpha_V=3.0$ and consider three different values for the number of non-resonant units, $N_{nr}=10$, $N_{nr}=30$, and $N_{nr}=70$, respectively. With this, in each one of these figures, we observe the formation of tongue-like structures of resonance across the ($D$,$c$) parameter space. For a fixed value of $c$, the onset of SR follows the traditional dependence with $D$ in which SR is optimal for a small range of $D$. Remarkably, for a fixed noise intensity $D$, the tilted configuration of the resonance tongue implies the existence of an interval of the coupling intensity $c$ for the existence of SR. Interestingly, within this $c$-interval there is an optimal (darker shade) and a maximum value for SR. This feature was previously reported for complex networks of homogeneous resonant units \cite{Gao2001}. However, in the presence of failing oscillators (heterogeneities), the optimum value of $c$ specifies the coupling intensity in which the resonant units are able to induce resonant motion in the non-resonant ones. As the number of non-resonant units increases from Fig. \ref{figure3}(a) to Fig. \ref{figure3}(c), a larger noise intensity is required for the onset of SR, and the overall value of $\eta$ decreases, signaling a weaker amplification of the periodic signal. Furthermore, for increasing $N_{nr}$, the tongue-like structure containing the parameters corresponding to the onset of SR gets narrower, indicating the necessity of more specific values of the pair ($D$,$c$). The ($D$,$c$)-parameter spaces for the scale-free topology ($BA_{m=2}$) with different values of $N_{nr}$ are shown in Figs. \ref{figure3}(d) - Figs. \ref{figure3}(f). The results are in accordance with the random topology.

\begin{figure}[!htp]
\centering
\includegraphics[width=85mm]{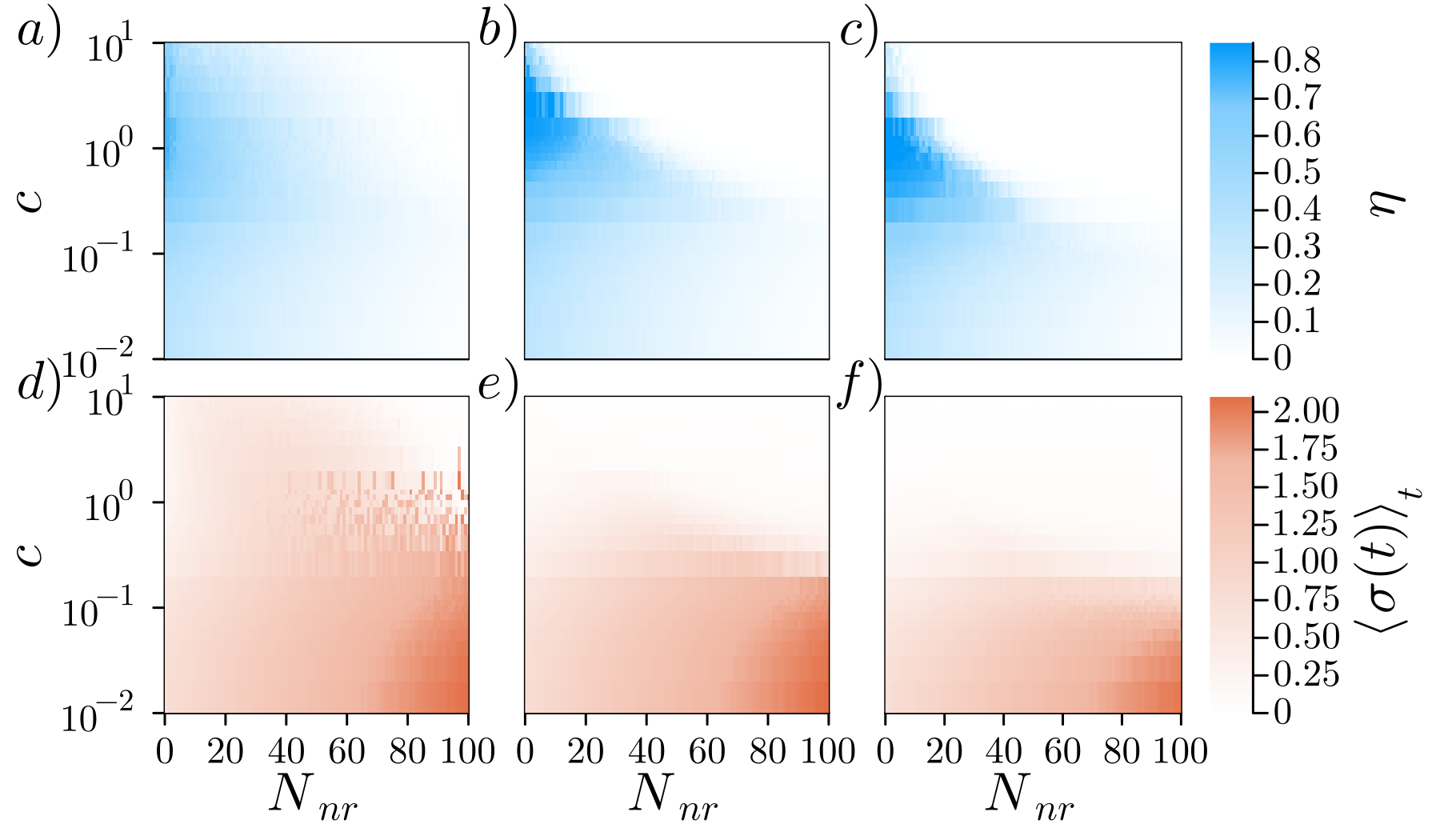}% Here is how to import EPS art
\caption{(a)-(c) Spectral amplification $\eta$ (color code) as a function of the coupling strength $c$ and the number of non-resonant oscillators $N_{nr}$ for the topologies $WS_{p=0}$, $WS_{p=1.0}$, and $BA_{m=2}$, respectively. (d)-(f) Mean square deviation $\langle \sigma(t)  \rangle_t$ as function of $c$ and $N_{nr}$. The topologies are the same as (a)-(c). The noise intensity is fixed at $D=1.0$.}
\label{figure4}
\end{figure}

Beyond the dependence of stochastic resonance on the network coupling and noise intensity, the analysis presented in Fig. \ref{figure3} already hints at the dependence on the number of non-resonant units. We now address this issue with special attention to the underlying role of the coupling intensity $c$. For that, we fix the noise intensity at $D=1.0$ and keep the dissimilarity index at $\alpha_V=3.0$. With this, in Fig. \ref{figure4}, we study the combined influence of $N_{nr}$ and $c$ in corresponding two-dimensional parameter space. From Figs. \ref{figure4}(a) to \ref{figure4}(c), we consider the three different topologies utilized along this study, regular ($WS_{p=0.0}$), random ($WS_{p=1.0}$), and scale-free ($BA_{m=2}$), respectively. For the regular topology in Fig. \ref{figure4}(a), we observe that the stochastic resonance phenomenon is very sensitive to the number of nonresonant units. Nevertheless, the occurrence of optimal values of $c$ for $N_{nr}$ slightly larger than zero is visible. On the other hand, for the random topology in Fig. \ref{figure4}(b), SR is more robust occurring for significantly larger values of $N_{nr}$. For this case, it is clear the occurrence of an optimal value of $c$ for the onset of SR. Therefore, the coupling intensity $c$ plays an essential role in regulating the interplay between the resonant and non-resonant units. A too-small value of $c$ is not enough to amplify the periodic input, while a too-large value favors the dynamics of non-resonant units breaking SR down. Similarly, for a scale-free topology in Fig. \ref{figure4}(c), an optimal value of $c$ is observed, indicating that SR in networks with failing oscillators is favored in complex topologies.

Similarly to the cases for the homogeneous network in the literature, the long-range connections of complex topologies favor SR as much as synchronization \cite{Gao2001}. To verify how this interplay responds to the failing oscillators introduced here, we assess the level of spatial synchronization in our network via the mean square deviation defined as \cite{Gao2001}:
\begin{eqnarray}
\sigma(t) &=& \frac{1}{N} \sum_{i=1}^{N} x_i(t)^2 - \left[ \frac{1}{N} \sum_{i=1}^{N} x_i(t) \right]^2.
\label{network_msd}
\end{eqnarray}
At a given time instant $t$, small values of $\sigma(t)$ indicate a high level of synchronization among the units, while high values of $\sigma(t)$ correspond to incoherent spatial dynamics. Since this function is oscillating, we obtain a time averaged value as $\langle \sigma(t)  \rangle_t = \frac{1}{T} \int_0^T \sigma(t) dt$.  With this, in Figs. \ref{figure4}(d) - \ref{figure4}(f), we obtain the values of $\langle \sigma(t)  \rangle_t$ for the corresponding parameter pairs ($N_{nr}$, $c$) for which the values of $\eta$ is shown in Figs. \ref{figure4}(a) - \ref{figure4}(c). By comparing the evaluations of $\eta$ and $\langle \sigma(t)  \rangle_t$, we verify that stochastic resonance correlates with synchronization in the network. Moreover, the increasing number of non-resonant units is equally detrimental to both SR and synchronization.

\begin{figure}[!htp]
\centering
\includegraphics[width=56.6mm]{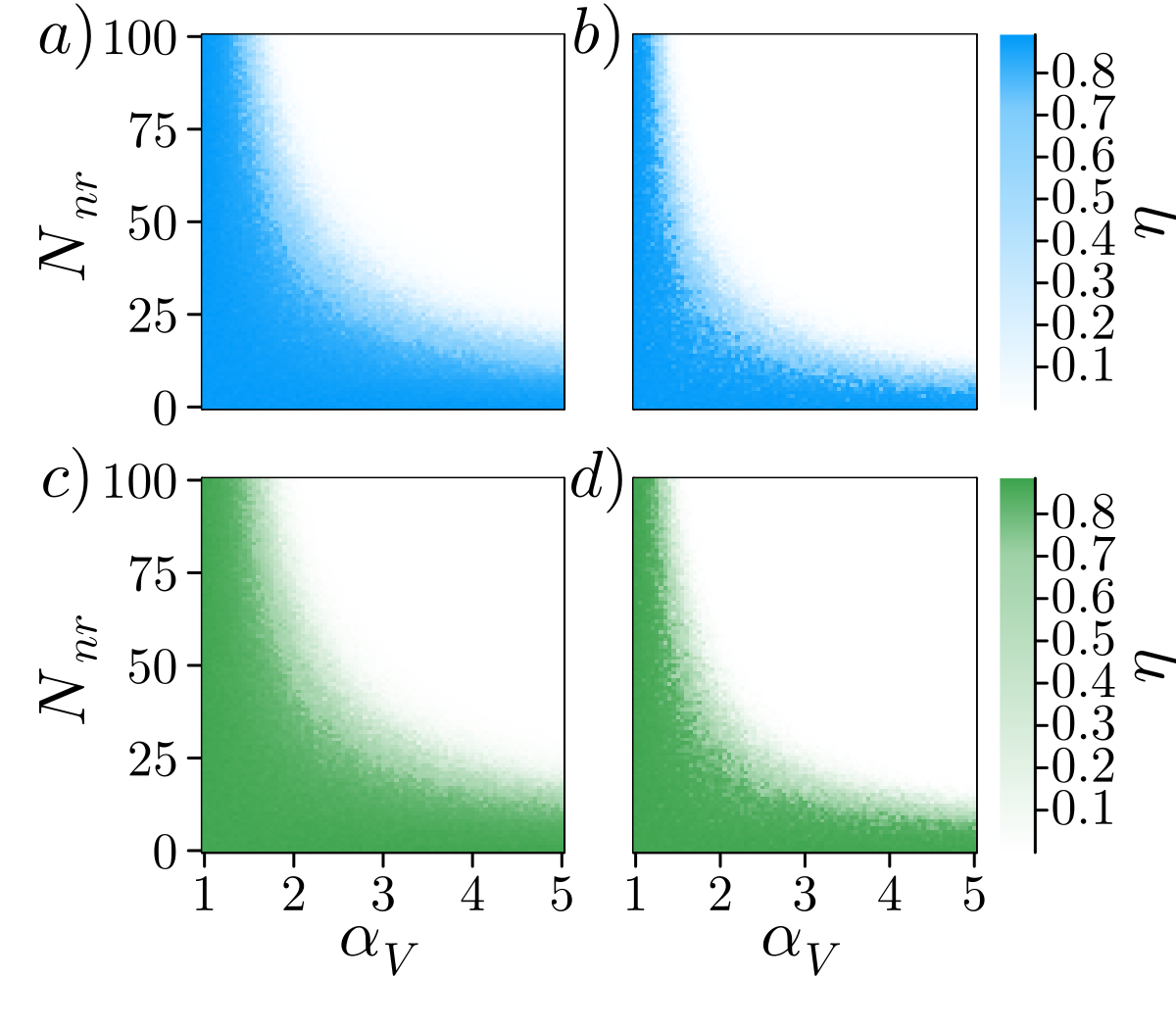}% Here is how to import EPS art
\caption{Spectral amplification $\eta$ (color code) as a function of the number of non-resonant units $N_{nr}$ and the dissimilarity index $\alpha_V$. (a)-(b) For a $WS_{p=0}$ topology, the coupling intensities are $c=1.0$ and $c=2.0$, respectively. (c)-(d) For a $BA_{m=2}$ topology. The noise intensity is fixed at $D=1.0$.}
\label{figure5}
\end{figure}

Subsequently, we unravel the relationship between the number of non-resonant units and their respective level of dissimilarity with the resonant units for which the network can still feature SR. For that, we first obtain the spectral amplification $\eta$ as a function of the number of non-resonant units $N_{nr}$ and their respective dissimilarity index. Hence, in Fig. \ref{figure5}, we color code the values of $\eta$ for $N_{nr} \in [0,N]$ and $\alpha_V \in [1,5]$. For the random network ($WS_{p=1.0}$) with coupling intensity $c=1.0$, in Fig. \ref{figure5}(a), the high values of $\eta$ (blue color) delimitate the area for the occurrence of stochastic resonance on the $\alpha_V,N_{nr}$ plane. For these values, the absence of inter-well dynamics in a subset of units, and their corresponding $\alpha_V$, is compensated by the resonant units via the action of the network coupling. On the other hand, for the values of $\eta$ approaching zero (white color), the resonant units are not able to establish stochastic resonance for the corresponding number of non-resonant units $N_{nr}$ and/or dissimilarity index $\alpha_V$. For such a region of ($\alpha_V,N_{nr}$)-plane, the stochastic resonance breaks down. In addition, there is a transition zone separating the resonant from the non-resonant network dynamics. At this zone, the network dynamics exhibit stochastic resonance intermittently in time. Furthermore, notice in Fig. \ref{figure5}(a) the nonlinear dependence of the number of non-resonant units $N_{nr}$ on the potential dissimilarity $\alpha_V$ for the onset of stochastic resonance. In Fig. \ref{figure5}(b), for a larger coupling intensity $c=2.0$, we verify that the parameter area leading to stochastic resonance has shrunk. This confirms the role of the coupling intensity in suppressing SR. To confirm these findings, for the scale-free (BA) topology, in Fig. \ref{figure5}(c) and in Fig. \ref{figure5}(d), we obtain $\eta(\alpha_V,N_{nr})$ for $c=1.0$ and $c=2.0$, respectively. Both, the nonlinear dependence of $N_{nr}$ on $\alpha_V$ and the shrinking of the range for SR are confirmed for this topology.

Finally, we inspect the characteristics of the transition zone between the resonant and non-resonant combinations of $N_{nr}$ and $\alpha_V$. As we observe in Fig. \ref{figure5}, the transitions for a random $WS_{p=1.0}$ and scale-free $BA_{m=2}$ topology are very much alike. We now investigate the influence of different rewiring probability $p$ in the Watts-Strogatz algorithm and different amounts of edges $m$ added at each step of the Barab\'{a}si-Albert algorithm. For that, we fix the dissimilarity index at $\alpha_V=3.0$ and investigate the transitions by varying only the number of non-resonant units $N_{nr}$.

\begin{figure}[!htp]
\centering
\includegraphics[width=85mm]{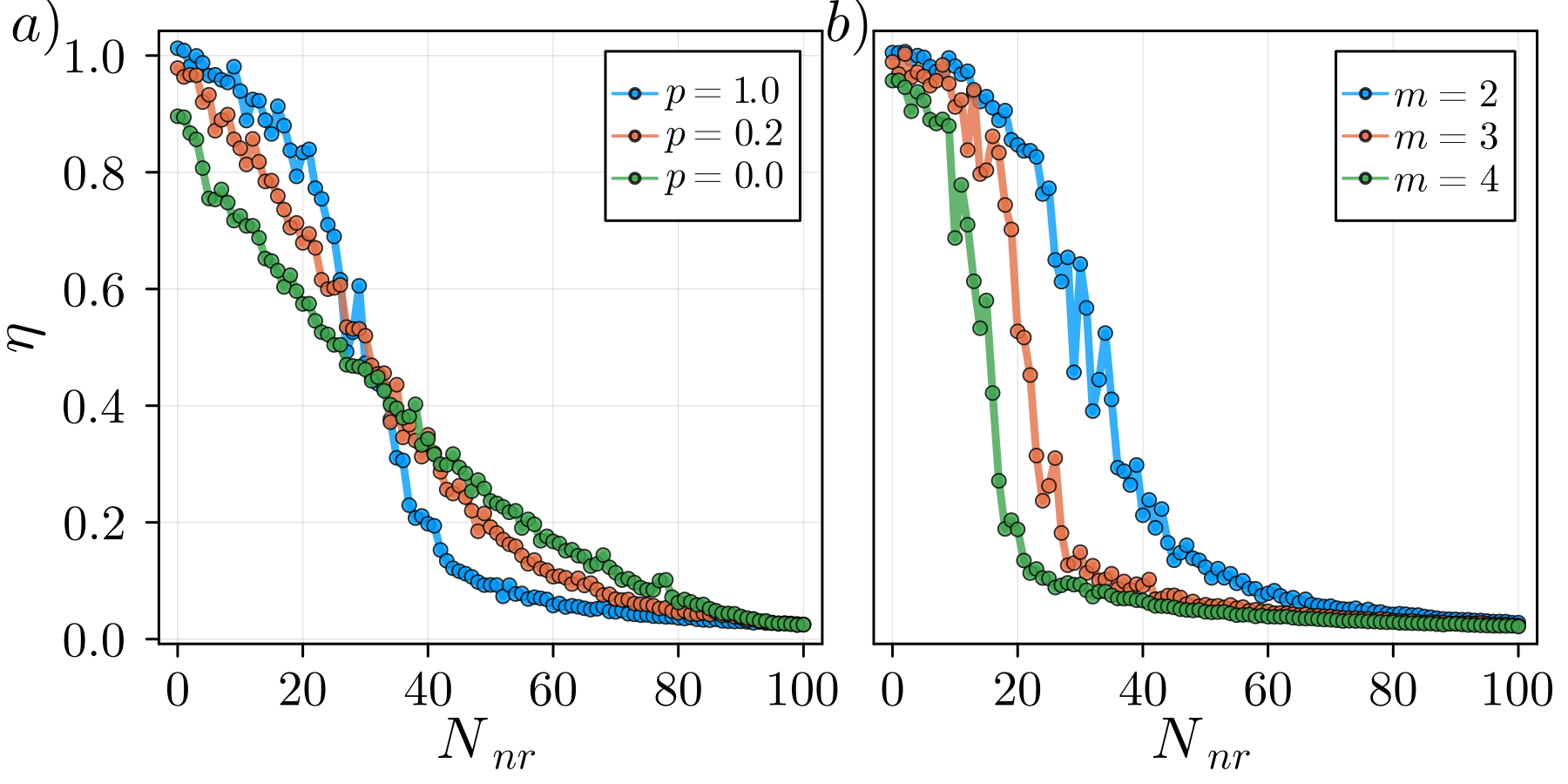}% Here is how to import EPS art
\caption{Spectral amplification $\eta$ as a function of the number of non-resonant units $N_{nr}$ for different topologies. a) For $WS_{p=0}$ (green), $WS_{p=0.2}$ (orange), and $WS_{p=1.0}$ (blue). (b) For $BA_{m=2}$ (blue), $BA_{m=3}$ (orange), and $BA_{m=2}$ (green). The coupling and the noise intensity are fixed at $c=1.0$ and $D=1.0$, respectively.}
\label{figure6}
\end{figure}

In Fig. \ref{figure6}(a), we examine a Watts-Strogatz topology with three different rewiring probabilities:  $p=0.0$ (green), $p=0.2$ (orange), and $p=1.0$ (blue). For $p=0.0$, yielding a regular nonlocal network, as we have seen before, for this topology SR is very sensitive to the increase of $N_{nr}$. The decrease in $\eta$ is smooth and uniform during the transition, a behavior attributed to the high clustering coefficient and path length typical of regular networks. For $p=0.2$, corresponding to a small-world topology with high clustering and low average path length, the spectral amplification is slightly higher in comparison with the regular case. Nevertheless, the characteristics of the transition to the non-resonant case are still smooth and uniform. In contrast, for a random topology obtained by setting $p=1.0$, the transition to the non-resonant parameter region is threshold-like. It means that despite $\eta$ constantly decreasing, there is an interval of $N_{nr}$ in which $\eta$ decreases faster (see the blue dots in Fig. \ref{figure6}(a)). This topology, with a similar average path length to the case with $p=0.2$ but a much lower clustering coefficient, provides evidence that stochastic resonance is better supported in topologies with low clustering coefficients.

In Fig. \ref{figure6}(b), we examine the Barab\'{a}si-Albert (BA) topology with three different values of the parameter $m$: $m=2$ (blue), $m=3$ (orange), and $m=4$ (green). Although the clustering coefficient of the BA topology is generally higher than a random one, we observe a threshold-like transition for all three considered values of $m$. However, the stochastic resonance is more persistent for lower values of $m$. We recall that higher values of $m$ imply increased connectivity in the network. Similar to the effect of increasing the coupling intensity (Fig~\ref{figure4}), higher connectivity causes SR to fail. Since the maintenance of SR in heterogeneous networks with failing oscillators relies on the resonant units overcoming the non-resonant ones, the process appears to be facilitated in sparsely connected networks. Beyond this interpretation, we highlight that the increasing abruptness of the breakdown of SR as the parameter $m$ increases is intriguing and warrants future investigation in the context of tipping points of oscillatory solutions \cite{Medeiros2017,Bathiany2018}.

\section{Low-dimension deterministic system}

\begin{figure}[!htp]
\centering
\includegraphics[width=85mm]{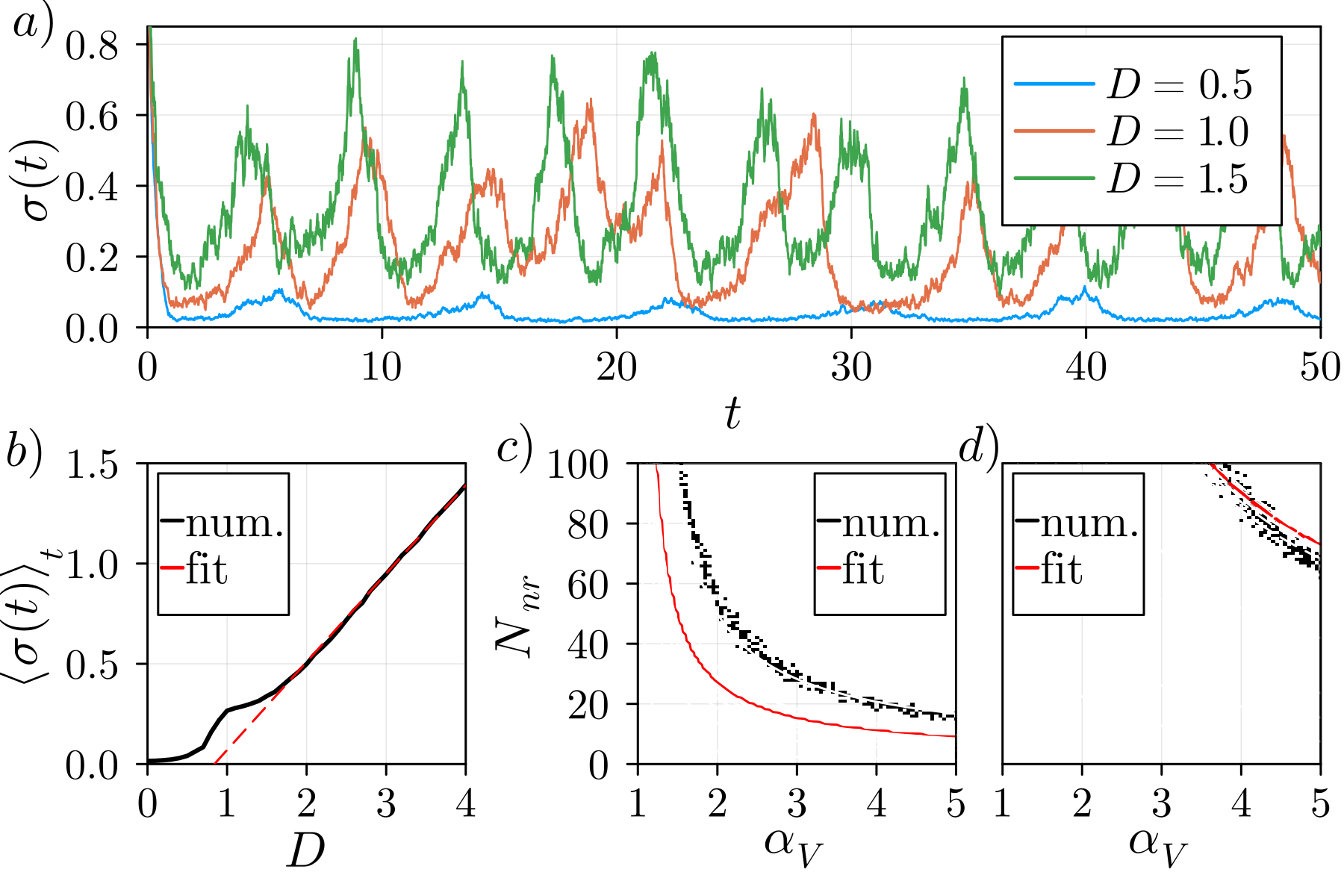}
\caption{(a) Time evolution of the mean square deviation $\sigma(t)$ given by Eq. (\ref{network_msd}) obtained for the network (Eq. (\ref{network_dgl_homo})) with random topology ($WS_{p=1.0}$). The heterogeneities prescribed as $N_{nr}=30$ and $\alpha_V=3.0$, while the coupling intensity is $c=1.0$. The colors mark different noise intensities as shown. (b) In black is the time-averaged amplitude $\langle \sigma(t)\rangle_t$ as function of $D$. The red dashed line stands for a linear fitting corresponding to the amplitude of $M(D,t)$ in Eq. (\ref{linampapprox}). (c) Parameter space ($\alpha_V$, $N_{nr}$) obtained for $D=1.0$. In black is the threshold obtained for the complete network, while red shows the thresholds obtained for the low-dimensional deterministic model. (d) Same as (c) for $D=2.0$.}
\label{figure7}
\end{figure}

So far, our results refer to a stochastic high-dimensional system, we now make considerations in order to reproduce these results in a low-dimensional deterministic system. To achieve this, we use the framework proposed in Refs. \cite{Wu2011,Tessone2006} to investigate the time evolution of the spatial mean variable $X(t) = \frac{1}{N} \sum_{i=1}^{N} x_i(t)$. The following symmetry arguments are used to reduce the system's dimensions through the averaging procedure: i) The coupling matrix $\textbf{M}$ is symmetric. Averaging over the units $i$ results in a null coupling term, i.e. $\sum_{i=1}^{N} \sum_{j=1}^{N} M_{ij}(x_j - x_i) = 0$. ii) The noise input $\xi_i$ is drawn from an even distribution, resulting in a null spatial mean, i.e. $\frac{1}{N}\sum_{i=1}^{N}\xi_i(t)=0$, which holds for a sufficiently large system \cite{Pikovsky2002}. iii) The periodic signal is independent of $i$ and is unaffected by the averaging procedure. Applying this procedure to Eq. (\ref{network_dgl_homo}) and considering the two subsets of the parameters $a_i$, the averaged equation of motion becomes:
\begin{eqnarray}
\nonumber
\frac{1}{N}\sum_{i=1}^{N}\dot{x}_i &=& \frac{1}{N}a\left(\sqrt{\alpha_V}\sum_{i=1}^{N_{nr}}x_i + \sum_{i=N_{nr}+1}^{N} x_i \right) + \\
&-&  \frac{1}{N}b\sum_{i=1}^{N}x_{i}^{3} + A\sin(\omega t).
\label{average_1}
\end{eqnarray}
We define a mean variable for each interval of $a_i$, i.e., $X_{nr} = \frac{1}{N_{nr}} \sum_{i=1}^{N_{nr}} x_i$ for the interval in which $a_i=\sqrt{\alpha_V}a$ (non-resonant) and $X_{r}= \frac{1}{(N-N_{nr})} \sum_{i=N_{nr+1}}^{N} x_i $ for the interval in which $a_i=a$ (resonant). However, as shown in Fig. \ref{figure2}(a), the position of the potential minimum is very similar for both resonant and non-resonant cases. Therefore, we assume that the average position in each potential is approximately equal to the global average, i.e., $X_{r} \approx X_{nr} \approx X$. In addition, to handle the cubic term in Eq. (\ref{average_1}), we express the position $x_i$ of an oscillator $i$ as a deviation of the average value, $x_i=X+\delta_i$. Hence, assuming that the deviations $\delta_i$ are Gaussian random variables with vanishing mean \cite{Tessone2006, Wu2011}, we obtain:
\begin{eqnarray}
\nonumber
 \dot{X} &=& \left[a \left( 1 - (1 - \sqrt{\alpha_V} )\frac{N_{nr}}{N} \right) - 3bM(t)\right]X + \\
 &-& bX^3 + A\sin(\omega t),
 \label{average_2}
\end{eqnarray}
where $M(t)=\frac{1}{N} \sum_{i=1}^{N} \delta^2_i$ stands for the variance of $X$. Thus, Eq. (\ref{average_2}) provides a deterministic low-dimensional approximation of our system. However, the time-dependent variance $M(t)$ remains unknown. To gather insights about the dynamics of $M(t)$, we resort to the high-dimensional system, where the mean square deviation $\sigma(t)$ is obtained using Eq. (\ref{network_msd}). In Fig. \ref{figure7}(a), we depict the time evolution of $\sigma(t)$ for three different noise intensities, $D=0.5$ (blue curve), $D=1.0$ (orange curve), and $D=1.5$ (green curve). We observe that the dynamical behavior of $\sigma(t)$ is strongly dependent on $D$, with its amplitude increasing as $D$ grows. To further demonstrate this behavior, Fig. \ref{figure7}(b) shows the time average $\langle \sigma(t) \rangle_t$ as function of $D$ (black curve), highlighting a linear relationship for $D \gtrsim 1.5$. Additionally, as $D$ increases, $M(t)$ exhibits a degree of regularity at the frequency $2\omega$. In light of these observations, we propose an {\it ansatz} to capture the linear growth and periodicity of $\sigma(t)$, which will be used as the variance $M(t)$ in our low-dimensional deterministic model described by Eq. (\ref{average_2}). Hence,
\begin{eqnarray}
    M(D,t) = \left( \alpha D + \beta \right)sin^2(\omega t),
    \label{linampapprox}
\end{eqnarray}
where $\alpha$ and $\beta$ are the parameters of the linear model for the amplitude, while the oscillations at frequency $2\omega$ are specified by the squared sine function with frequency $\omega$. Next, we fit the linear model for the amplitude to the network data in Fig. \ref{figure7}(b) (red dashed line). The fitting procedure yields the following values for the model parameters, $\alpha = 0.44 \pm 0.02$ and $\beta = 0.37 \pm 0.06$. Finally, we replace the model (Eq. (\ref{linampapprox})) with the corresponding parameters in the reduced system in Eq. (\ref{average_2}). This allows us to estimate the spectral amplification $\eta$ of the reduced system. Hence, for $D=1.0$, in the parameter plane ($\alpha_V$, $N_{nr}$) shown in Fig. \ref{figure7}(c), we compare the thresholds obtained for the complete system (black curve) with the low-dimensional deterministic model (red dashed line). We verify that the threshold prescribed by the model differs significantly from the one observed for the network. We attribute this deviation to the poor fitting of the amplitude for lower values of $D$ shown in Fig. \ref{figure7}(b). Conversely, for higher values of $D$, for instance, $D=2.0$ shown in Fig. \ref{figure7}(d), the low-dimensional model provides a better reproduction of the network threshold, thereby validating our analysis.

\section{Conclusion}

In conclusion, we have investigated the breakdown of the stochastic resonance (SR) phenomenon in networks of bistable oscillators. We perturbed the functioning of the resonant network by introducing failing units possessing non-resonant potentials (heterogeneities). First, we found that the combinations of network coupling and noise intensity for the maintenance of SR form tongue-like structures, specifying different optimal values of these parameters depending on the number of non-resonant units. Subsequently, we verified that, counter-intuitively, increasing the network coupling intensity causes SR to globally fail in the presence of a few non-resonant oscillators. Additionally, the upper limit of the coupling intensity diminishes as the number of non-resonant units increases. Moreover, we observed that for the maintenance of SR in our network, the number of non-resonant oscillators is a nonlinear function of their dissimilarity level.

Considering different network structures, we report that for a regular topology, SR is very sensitive to the introduction of failing units, quickly vanishing as the number of non-resonant units increases. Similarly, for a complex topology with a high clustering coefficient and low average path length, the sustainability of SR with an increasing number of non-resonant units is only marginally better than in the regular case. However, for a random topology, featuring both a low clustering coefficient and average path length, the network is able to compensate for a large number of non-resonant units, better sustaining SR. For a scale-free topology, we found the persistence of SR to be related to the connectivity of the topology. Specifically, we observe that a sparse scale-free network offers better support for this phenomenon.

Among the possible applications of our findings, we highlight the potential for using SR as a diagnostic method to detect failing units in complex networks. Specifically, since our results specify the weakening of SR as a function of the number of failing units for a given network coupling intensity, it enables the verification of the number of failing units in the network once the intensity of SR for the normal network is known. Additionally, these results establish the tolerance levels for the number of failing units so that SR can still occur. These strategies can be useful for ensuring the robustness of SR in networks with various purposes, such as lasers \cite{McNamara1988, Barbay2000}, electronic circuits \cite{Anishchenko1992, Anishchenko1994, Mantegna1994}, and optomechanical systems \cite{Monifi2016}. Moreover, since SR has been found to improve the prediction accuracy and stability of machine learning algorithms \cite{Zhai2023}, our results assist in designing more robust artificial neural networks that can benefit from the SR phenomenon.

\begin{acknowledgments}
E.S.M acknowledges support from The S\~ao Paulo Research Foundation (FAPESP), project number 2023/15040-0, and the Deutsche Forschungsgemeinschaft (DFG), project number 454054251. This work was supported by the Deutsche Forschungsgemeinschaft (DFG, German Research Foundation) - project number 163436311 - SFB 910.
\end{acknowledgments}

\bibliographystyle{unsrt}

% \bibliography{Bibliography.bib}% Produces the bibliography via BibTeX.

\end{document}